\documentclass[a4paper, 12pt, openany, oneside]{article}
\usepackage[cp1251]{inputenc} %- для WiN
\usepackage[english, russian]{babel} %- для рус-англ. переноса
\usepackage{amsmath, latexsym,  amssymb, bm,  graphics, eucal,
amsfonts, array}
\usepackage[dvips]{graphicx}
\usepackage{epstopdf}
\makeatletter
\renewcommand{\@biblabel}[1]{#1.\hfill}

\makeatother
\usepackage{indentfirst}
\makeatother

\begin{document}
\renewcommand{\refname}{\begin{center} \bf References\end{center}}
 \thispagestyle{empty}
\renewcommand{\abstractname}{Abstract}
\renewcommand{\contentsname}{Contents}
 \large

 \begin{center}
\bf Solution of the Kramers' problem about isothermal sliding of
moderately dense gas with accomodation boundary conditions
\end{center}\bigskip

\begin{center}
\copyright\quad {\bf2011 г.\quad A. V. Latyshev, A. D. Kurilov}%А. А. Юшканов}
\end{center}\medskip \bigskip

\begin{abstract}
Half-space boundary  Kramers' problem about isothermal sliding of moderate dense gas with accommodation boundary conditions along a flat firm surface is solving. The new method of the solution of boundary problems of the kinetic theory is applied (see JVMMF, 2012, {\bf 52}:3, 539-552).
The method allows to receive the solution with arbitrary degree of accuracy. The idea of representation of boundary condition on distribution function in the form of source in the kinetic equation serves as the basis for the method mentioned above. By means of Fourier integrals the kinetic equation with a source comes to the Fredholm integral equation of the second kind. The solution has been received in the form of Neumann's number.

\end{abstract}
\tableofcontents
\setcounter{secnumdepth}{4}

\begin{center}
\item{}\section {Introduction}
\end{center}

The Kramers' problem is one of the major problems
in kinetic theory of gases. This problem has great practical importance.
The solution of this problem is set forth in such monographies, as
\cite {Ferziger} and \cite {Cerc73}.

55 years ago precisely K. M. Case in the well-known work
\cite {Case60} laid the foundation of the analytical solution
of boundary problems of the transport theory.
The idea of this method included the following: to find the general
solition of the non-uniform characteristic equation corresponding
to the trans\-port equation, in class of the generalized functions
in the form of the sum two generalized functions (
an integral principal value and composed, proportional to
Dirac delta function).

The first of those сomposed is the partial solution of non-uniform
charac\-te\-ris\-tic equation, and the second is the common solution
of corresponding homoge\-neous equation answering
to the non-uniform characteristic equation.
As proportionality coefficient in this expression costs so
named dispersion function. Zero of dispersion function
are connected biunique with partial solution of
the initial transport equations.

We come to the  characteristic equation 
after division of variables 
in the transport equation. By means of spectral parametre we
divide spatial and velocity variables in the transport equation.

The general solution of the characteristic equation include as a 
particular solution singular Cauchy kernel,
which denominator is a difference of velocity and spectral
variable.

Cauchy kernel provides the use of all powers of methods of the theory
functions complex variable, in particular,  theory of boundary
value Riemann---Hilbert problems.

So, construction of eigen functions of the characteristic
equations leads to concept of the dispersion equation, which 
roots are in biunique conformity with
partial (discrete) solutions of the initial transport equation.

The general solution of boundary problems for the transport equation
is searched in form of  linear combination of discrete solutions with
arbitrary
coefficients (these coefficients are called as discrete coefficients)
and integral on spectral parametre from the eigen
functions of the continuous spectrum, multiplicated by unknown function
(coefficient of continuous spectrum).
Some discrete coefficients are setting and considered
as known. Discrete coefficients answer to the discrete
spectrum, or, in some cases, answer to spectrum,
attached to the continuous.

Substitution of general solution in boundary conditions leads
to singular integral equation with Cauchy kernel. The solution of it
allows to construct the solution of an initial boundary problem of
the transport equations.

Acting in such a way, Carlo Cercignani in 1962 in work
\cite{Cerc62} const\-ruc\-ted the exact solution of Kramers' problem
with pure diffusion boundary conditions about isothermal sliding.
Cercignani in \cite{Cerc62}
gave the exact solution of linear kinetic stationary Boltzmann equation
with integral of collisions in form of $\tau$--model.
Here with there were used diffusion boundary conditions.

Works \cite{Case60, Cerc62} laid down the foundations of the analytical
methods for finding of exact solutions of the modelling kinetic
equations.

Generalization of this method on a vector case (system of the kinetic equations)
with constant frequency of collisions encounters on the considerable
difficulties (see, for example, \cite{Siewert74}).
Such difficulties have faced authors of works \cite{Siewert74, Cerc77}.
To overcome these difficulties it was possible in work \cite{LatPMM}, in which
the problem about temperature jump has been solved.

The problem about temperature jump with frequency
of collisions, propor\-tional to the module of molecule velocity,
it has been solved by Wiener---Hopf's method in work \cite{Cassell}.
Then in more general statement taking into account weak evaporation
(condensation) this problem has been solved by Case' method in our
work \cite{Mgg96}. These problems are reduced to the solution of the
vector kinetic equations.

The Kramers' problem has been generalized further on the case of the binary
gases \cite{TMF91} and \cite{JVMMF91},
it has been solved with use of the higher models
of Boltzmann equations \cite{Model1}, it has been generalized on the case
accomodation \cite{Boundary} and mirror--diffusion boundary
conditions \cite{Boundary1}.

Last decade the Kramers' problem has been formulated and
it is analyti\-cally solved for quantum gases \cite{Fermi1} and \cite{Bose1}.

In our works \cite{Method1}, \cite{Method2} and \cite{method2012}
the approximation methods of the solution of boundary problems
of kinetic theory have been developed.

In the present work we use a new effective method of the solution
of boundary problems with mirror--diffusion boundary conditions
\cite{method2012} for the solution of Kramers' problem for moderately dense gas.

The question on integral of collisions for dense gases has been 
studying during last decades \cite{Res}. In the solution of this 
very question there are various approaches, since Enskog'
models \cite{Fer} and finishing modern decomposition of integral
collisions abreast on density degrees. General line,
characteristic for all app\-ro\-aches, is nonlocal property of integral
collisions for dense gases.

In work \cite{Past1} the  model of integral collisions for moderately
dense gas has been offered. Then in works
\cite{Past2} -- \cite{Gaj} this model has been used in
concrete problems.
Besides, in work \cite{Lat94} the Kramers' problem for moderately dense
gas with diffusion boundary conditionsis has been
formulated and analytically solved.

At the heart of an offered method  lays the idea to include the boundary
condition on function of distribution in the form of a source in the kinetic
equation.

The main point of an offered method include the following. Initially 
our Kramers' problem for moderately dense gas about isothermal sliding with
mirror--diffusion boundary conditionsis formulated in half-space $x>0$.
Then distribution function proceeds in
the conjugated (interfaced)  half-space $x <0$ in the even manner on to
spatial and on velocity variables. In half-space
$x <0$ also are formulated an initial problem.

After it is received the linear kinetic equation
let us break required function (which we will name also
distribution function) on two composed: Chapman---Enskog'
distribution function $h_{as}(x, \mu) $ and the second part of
distributions function $h_c(x,\mu) $, correspond to the continuous spectrum
$$
h (x, \mu) =h _ {as} (x, \mu) +h_c (x, \mu).
$$
\noindent($as \equiv asymptotic, c\equiv continuous $).

By virtue of the fact that The Chapman—Enskog distribution function 
is a linear combination of discrete solutions of the initial equation,
the function $h_c(x,\mu) $ also is the solution of the kinetic
equations. Function $h_c(x, \mu) $ passes in zero far from
wall. On a wall this function satisfies to mirror--diffusion boundary condition.

Further we will transform the kinetic equation for function
$h_c(x, \mu)$. We include in this equation the boundary condition on
to wall for function $h_c (x, \mu)$ in the form of a member of type
the source laying in a plane $x=0$. We will underline, that function
$h_c(x,\mu) $ satisfies to the received kinetic equation in
both conjugated  half-spaces $x <0$ and $x> 0$.

We solve this kinetic equation in the second and the fourth
quadrants of phase plane $(x,\mu) $ as the linear
differential equation of the first order, considering known
mass velocity of gas $U_c(x)$. From the received solutions
it is found the boundary values of unknown function $h^{\pm}(x,\mu)$
at $x =\pm 0$, entering into the kinetic equation.

Now we expand unknown function $h_c(x,\mu)$, unknown mass velocity
$U_c(x)$ and delta-Dirac function as Fourier integrals.
Boundary values of unknown function $h_c^{\pm}(0,\mu)$
are thus expressed by the same integral on the spectral
density $E(k)$ of mass velocity.

Substitution of Fourier integrals in the kinetic equation and
expression of mass velosity leads to the characteristic
system of the equations. If to exclude from this system the spectral
density $\Phi(k,\mu)$  of function $h_c(x, \mu) $, we will receive
Fredlolm integral equation of the second kind.

Considering the gradient of mass speed is set, we will expand
the unknown velocity of sliding, and also spectral density of mass
velocity and distribution functions in series by degrees on
diffusions coefficients $q $ (these are Neumann's numbers).
On this way we receive numerable
system of the hooked equations on coefficients of series for
spectral density. Thus all equations on coefficients of
spectral density for mass velocity have singularity (a pole
of second order in zero). Excepting these singularities
consistently, we will construct all members of  series for
velocity slidings, and also series for spectral density of the mass
velocity and distribution function.

\begin{center}
\item{}\section{Raising an issue}
\end{center}

Assuming that moderately dense gas occupies half-space $x>0$ over firm flat
mo\-ti\-on\-less wall. We take the Cartesian system of coordinates with an axis $x $,
per\-pen\-di\-cu\-lar wall, and with plane ($y, z$), coinciding with a wall, so
that the origin of coordinates lays on a wall.

Let us assume, that far from a wall and along an axis $y$
the gradient of mass velocity of the gas which quantity is equal $g_v $ is set.
The setting of gradient of mass velocity of gas causes
motion of gas along a wall.
Let us consider this motion in absence of the tangential
gradient of pressure and at to constant temperature.
In these conditions the mass velocity of gas will have
only one tangential component $u_y(x)$, which
far from a wall will vary under the linear law. A deviation from
linear distribution will be to occur near to a wall in a layer, often
named Knudsen' layer,
which thickness has an order of length of the mean free
path $l $. Out of Knudsen' layer the gas motion is described by the equations
of Navier---Stokes. The phenomenon of movement of gas along a surface, caused
the gradient of mass velocity set far from a wall, is called
isothermal sliding of gas.
For the solution of Navier---Stokes equations it is required to put
boundary conditions on to wall. As such boundary condition is accepted
the extrapolated value of hydrodynamic velocity on a surface
(quantity $u_{sl}$. We note that a real profile of velocity in
Knedsen' layer it is distinct from the hydrodynamic.
For reception quantity $u_{sl}$
it is required to solve  the Boltzmann equation in Knudsen' layer.

At the small gradients of velocity it is had
$$
u_{sl} =K_{v} l g_v, \qquad
g_v =\left (\dfrac {du_y(x)}{dx}\right)_{x\to + \infty}.
$$

Problem of finding of isothermal velocity sliding $u_{sl}$
is called as Kramers' problem (see, for example, \cite{Ferziger}).
Definition of the quantity  $u_{sl} $
allows, as we will see more low, completely construct the
distributionunction of gas molecules in the given problem.
And find the profile of distribution of mass velocity of gas
in half-space, and also find the quantity of mass velocity
of gas on  half-space border.

As the kinetic equation for distribution function
let us use the linear kinetic stationary Boltzmann equation
with integral of collisions for mode\-ra\-tely dense gas
$$
{v}_x\dfrac{\partial \varphi(x,{\bf C})}{\partial x}=
\nu\Big[2C_yu_y(x)+2\gamma C_yC_x\dfrac{du_y(x)}{dx}-\varphi(x,{\bf C})\Big].
\eqno{(2.1)}
$$
Here for molecules---hard spheres,
$$
\gamma=\dfrac{4}{15}\pi n \sigma^3,
$$
where $\sigma$ is the effective diameter of molecules, $n$ is the
concentration (number density) of gas.

In equation (2.1) $\nu$ is the effective collision frequency, $\tau=1/\nu$
is the сharacteristic time between two consecutive collisions,
${\bf C}$ is the dimensionless velocity of gas molecules,
$ \mathbf {C} = \sqrt{\beta}\mathbf {v} $,
function $ \varphi (x, {\bf C}) $ is connected with absolute
Maxwellian equality
$$
f(x,{\bf C})=f_0(v)[1+\varphi(x,{\bf C})],
$$
where $f_0(v)$ is the absolute Maxwellian,
$$
f_{0}(x,\mathbf{v})=n_0\Big(\dfrac{\beta}{\pi}\Big)^{3/2}\exp
\Big\{-\beta[v_x^2+v_y^2+v_z^2]\Big\},\qquad
\beta=\dfrac{m}{2kT},
$$
$\mathbf{v}=(v_x,v_y,v_z)$ is the velocity of molecule,
$k$ is the Boltzmann constant, $m$ is the mass of molecule,
$n$ and $T$ are accordingly numerical density (con\-cen\-tra\-tion) and temperature,
considered in the given problem are constants.

Let  us present the equation (1.1) in the dimensionless form.
For this purpose let us enter dimensionless mass velocity in an
axis direction $y $ $U (x) = \sqrt{\beta} u_y (x) $
and coordinate $x_1 =\nu \sqrt {\beta} x $. Clearly, that
dimensionless gradient of mass velocity
$$
G_v =\Big (\dfrac {dU} {dx_1} \Big) _ {x_1\to + \infty}
$$

It is connected with  gradient $g_v $ by equality $G_v=g_v/\nu
$.

Now the equation (2.1) registers in the form
$$
C_x\dfrac{\partial \varphi(x_1,{\bf  C})}{\partial x_1}
+\varphi(x_1,{\bf C})=
$$
$$
=\dfrac{2C_y}{\pi^{3/2}}\int e^{-C'^2}C_y'
\Big[\varphi(x_1,{\bf C'})+
\gamma C_x \dfrac{d\varphi(x_1,{\bf C'})}{dx_1}\Big]d^3C'.
\eqno{(2.2)}
$$

Let us continue function of distribution to the conjugated  half-space
in the symmetric manner
$$
f(x_1,\mathbf{C})=f(-x_1,-C_x,C_y,C_z).
\eqno{(2.3)}
$$

Continuation according to (2.3) on half-space
$x_1 <0$ allow to include boundary conditions in the problem equations.

Such continuation of function of distribution
allows to consider actually two problems, one of which
the second is defined in "positive"\, half-space $x_1> 0$, and in
"negative"\, half-space $x_1 <0$.

Let us formulate mirror--diffusion boundary conditions for
distribution function accordingly for "positive" \, and for
"negative" \, half-space
$$
f(+0, \mathbf{C})=qf_0(C)+(1-q)f(+0,-C_x,C_y, C_z), \quad C_x>0,
\eqno{(2.4)}
$$
$$
f(-0, \mathbf{C})=qf_0(C)+(1-q)f(-0, -C_x,C_y,C_z),\quad C_x<0.
\eqno{(2.5)}
$$

Here $q$ is the diffusion coefficient, $0 \leqslant q \leqslant 1$,
$f_0(C)$ is the absolute Maxwellian,
$$
f_0(C)=n\Big(\dfrac{\beta}{\pi}\Big)^{3/2}\exp(-C^2).
$$

In the equations (2.4) and (2.5) parametre $q $ is the
part of the molecules, dissipating  diffusion on border, 1$-q $
is the part of molecules dissipating mirror.

Let's search for distribution function in the form
$$
\varphi(x_1,{\bf C})=2C_y h(x_1,\mu).
$$

According (2.3) we have
$$
h(x_1,\mu)=h(-x_1,-\mu),\qquad \;\mu>0.
$$

The equation (2.2) transforms following manner
$$
C_x\dfrac{\partial h}{\partial x_1}+h(x_1,\mu)=
2U(x_1)+2\gamma \dfrac{dU(x_1)}{dx_1}.
$$

We obtain the formula for finding of mass velocity
$$
U(x_1)=
\dfrac{1}{\sqrt{\pi}}\int\limits_{-\infty}^{\infty}\exp(-C_x^2)h(x_1,C_x)
dC_x.
$$

Therefore, for function $h(x_1,\mu)$ we obtain the equation
$$
\mu\dfrac{\partial h}{\partial x_1}+h(x_1,\mu)=\dfrac{2}{\sqrt{\pi}}
\int\limits_{-\infty}^{\infty}e^{-t^2}\Big[h(x_1,t)+
\gamma \mu\dfrac{\partial h(x_1,t)}{\partial x_1}\Big]\,dt.
\eqno{(2.6)}
$$

Boundary condition (2.4) and (2.5) transforms into following boundary
condition
$$
h(+0,\mu)=(1-q)(+0,-\mu)=(1-q)(-0,\mu), \quad \mu>0,
\eqno{(2.4')}
$$
$$
h(-0,\mu)=(1-q)(-0,-\mu)=(1-q)(+0,\mu), \quad \mu<0.
\eqno{(2.5')}
$$

The right part of the equation (2.6) is the sum of mass velocity of gas
$$
U(x_1)=\dfrac{1}{\sqrt{\pi}}\int\limits_{-\infty}^{\infty}
e^{-t^2}h(x_1,t)dt
$$
and its derivative on coordinate.

Let us present function $h(x_1,\mu) $ in the form
$$
h(x_1,\mu)=h^{\pm}_{as}(x_1,\mu)+h_c(x_1,\mu),
\quad \text{если}\quad
\pm x_1>0,
$$
where asymptotic  part of distribution function (so-called
Chapman---Enskog distribution function)
$$
h_{as}^{\pm}(x_1,\mu)=U_{sl}(q)\pm G_v[x_1 -(1-\gamma)\mu],
\quad \text{если}\quad
\pm x_1>0,
\eqno{(2.7)}
$$
also is the solution of the kinetic equation (2.6).

Here $U_{sl} (q)$ is the required velocity  (dimensionless) of the isothermal
slidings.

As far from the wall ($x_1\to \pm \infty $) distribution function
$h(x_1,\mu)$ passes into Chapman---Enskog
$h_{as}^{\pm}(x_1, \mu) $, for function $h_c (x_1, \mu) $, correponding
to continuous spectrum, we receive the following boundary condition
$$
h_c(\pm \infty,\mu)=0.
$$

From here for mass velocity of gas it is received
$$
U_c(\pm \infty)=0.
\eqno{(2.8)}
$$

Let us notice, that in equality (2.7) sign on a gradient in "negative" \,
half-space varies on the opposite. Therefore the condition (2.8)
it is carried out automatically for functions $h_{as}^{\pm}(x_1,\mu)$.

Then boundary conditions $ (2.4 ') $ and $ (2.5') $ pass in the following
$$
h_c(+0,\mu)=$$$$=
-h_{as}^+(+0,\mu)+(1-q)h_{as}^+(+0,-\mu)+%$$$$+
(1-q)h_c(+0,-\mu),
\quad \mu>0,
$$
$$
h_c(-0,\mu)=$$$$=
-h_{as}^-(-0,\mu)+(1-q)h_{as}^-(-0,-\mu)+%+$$$$+
(1-q)h_c(-0,-\mu),  \quad \mu<0.
$$

We denote
$$
h_0^{\pm}(\mu)=-h_{as}^{\pm}(+0,\mu)+(1-q)h_{as}^{\pm}(+0,-\mu)=
$$
$$
=-qU_{sl}(q)+(2-q)(1-\gamma)G_v|\mu|.
$$

 Let us rewrite the previous boundary conditions in the form
$$
h_c(+0,\mu)=h_0^+(\mu)+(1-q)h_c(+0,-\mu), \quad \mu>0,
$$
$$
h_c(-0,\mu)=h_0^-(\mu)+(1-q)h_c(-0,-\mu), \quad \mu<0,
$$
where
$$
h_0^{\pm}(\mu)=-h_{as}^{\pm}(0,\mu)+(1-q)h_{as}^{\pm}(0,-\mu)=
$$
$$
=-qU_{sl}(q)+(2-q)(1-\gamma)G_v|\mu|.
$$

Considering symmetric continuation of function of distribution, we have
$$ h_c(-0,-\mu)=h_c(+0,+\mu),\qquad
h_c(+0,-\mu)=h_c(-0,+\mu).
$$
Hence, boundary conditions will be rewritten in the form
$$
h_c(+0,\mu)=h_0^+(\mu)+(1-q)h_c(-0,\mu), \quad \mu>0,
\eqno{(2.9)}
$$
$$
h_c(-0,\mu)=h_0^-(\mu)+(1-q)h_c(+0,\mu), \quad \mu<0.
\eqno{(2.10)}
$$

Let us include boundary conditions (2.9) and (2.10) in the kinetic
equation as follows
$$
\mu \dfrac{\partial h_c}{\partial x_1}+h_c(x_1,\mu)=2U_c(x_1)+
2\gamma \mu \dfrac{dU_c(x_1)}{dx_1}+
$$
$$
+|\mu|\Big[h_0^{\pm}(\mu)-q h_c(\mp 0,\mu)\Big]\delta(x_1),
\eqno{(2.11)}
$$
where $U_c(x_1)$ is the part of mass velocity corresponding to the continuous
spectrum,
$$
U_c(x_1)=\dfrac{1}{\sqrt{\pi}}\int\limits_{-\infty}^{\infty}
e^{-t^2}h_c(x_1,t)\,dt.
\eqno{(2.12)}
$$

Really, let, for example, $ \mu> 0$.
Let us integrate both parts of the equation
(2.11) on $x_1$ from $-\varepsilon $ to $ + \varepsilon $.
It is as result we receive equality
$$
h_c(+\varepsilon,\mu)-h_c(-\varepsilon,\mu)=h_0^+(\mu)-
qh_c(-\varepsilon,\mu),
$$
whence at $ \varepsilon\to 0$ in accuracy it is received the
boundary condition (2.9).

On the basis of definition of mass velocity (2.12) it is concluded, that
for it the condition (2.8) is satisfied
$$
U_c(+\infty)=0.
$$

Hence, in half-space $x_1> 0$  profile of mass velocity of gas
it is calculated under the formula
$$
U(x_1)=U_{as}(x_1)+\dfrac{1}{\sqrt{\pi}}\int\limits_{-\infty}^{\infty}
e^{-t^2}h_c(x_1,t)dt,
$$
and far from a wall has following linear distribution
$$
U_{as}(x_1)=U_{sl}(q)+G_vx_1, \qquad x_1\to +\infty.
$$

\begin{center}
\item{}\section{Kinetic equation in the second and the fourth
quadrants of phase plane}
\end{center}

Solving the equation (2.11) at $x_1> 0, \, \mu <0$, considering set the mass
velocity $U(x_1) $, we receive, satisfying boundary conditions (1.10),
the following solution
$$
h_c^+(x_1,\mu)=-\dfrac{1}{\mu}\exp(-\dfrac{x_1}{\mu})
\int\limits_{x_1}^{+\infty} \exp(+\dfrac{t}{\mu})2\Bigg[U_c(t)
+\gamma\mu \dfrac{dU_c(t)}{dt}\Bigg]\,dt.
\eqno{(3.1)}
$$

Similarly at $x_1 <0, \, \mu> 0$ it is found
$$
h_c^-(x_1,\mu)=\dfrac{1}{\mu}\exp(-\dfrac{x_1}{\mu})
\int\limits_{-\infty}^{x_1} \exp(+\dfrac{t}{\mu})2\Bigg[U_c(t)
+\gamma\mu \dfrac{dU_c(t)}{dt}\Bigg]\,dt.
\eqno{(3.2)}
$$

Now the equations (2.11) and (2.12) can be copied, having replaced
the second member into square bracket from (2.11) according to (3.1)
and (3.2), in the form
$$
\mu\dfrac{\partial h_c}{\partial x_1}+h_c(x_1,\mu)=$$$$=2U_c(x_1)+
2\gamma \mu \dfrac{dU_c(x_1)}{dx_1}+
|\mu|\Big[h_0^{\pm}(\mu)-qh_c^{\pm}(0,\mu)\Big]\delta(x_1),
\eqno{(3.3)}
$$
$$
U_c(x_1)=\dfrac{1}{\sqrt{\pi}}\int\limits_{-\infty}^{\infty}
e^{-t^2}h_c(x_1,t)dt.
\eqno{(3.4)}
$$

In equalities (3.3) boundary values $h_c^{\pm}(0,\mu)$
are expressed through the component of the mass velocity
corresponding to continuous spectrum
$$
h_{c}^{\pm}(0,\mu)=-\dfrac{1}{\mu}e^{-x_1/\mu} \int\limits_{0}^{\pm
\infty}e^{t/\mu}2\Bigg[U_c(t)
+\gamma\mu \dfrac{dU_c(t)}{dt}\Bigg]\,dt.
$$

For the solution of the equations (3.4) and (3.3) we search in the form
of Fourier integrals
$$
U_c(x_1)=\dfrac{1}{2\pi}\int\limits_{-\infty}^{\infty}
e^{ikx_1}E(k)\,dk,\qquad
\delta(x_1)=\dfrac{1}{2\pi}\int\limits_{-\infty}^{\infty}
e^{ikx_1}\,dk,
\eqno{(3.5)}
$$
$$
h_c(x_1,\mu)=\dfrac{1}{2\pi}\int\limits_{-\infty}^{\infty}
e^{ikx_1}\Phi(k,\mu)\,dk.
\eqno{(3.6)}
$$

Thus distribution function $h_c^+(x_1, \mu) $ is expressed through
spectral density $E (k) $ of mass velocity as follows
$$
h_c^+(x_1,\mu)=-\dfrac{1}{\mu}\exp(-\dfrac{x_1}{\mu})
\int\limits_{x_1}^{+\infty} \exp(+\dfrac{t}{\mu})dt
\dfrac{1}{2\pi}
\int\limits_{-\infty}^{+\infty}e^{ikt}E(k)(1+i\gamma\mu k)\,dk=
$$
$$
=\dfrac{1}{2\pi}\int\limits_{-\infty}^{\infty}e^{ikx_1}
\dfrac{1+i\gamma k \mu}{1+ik\mu}E(k)\,dk=$$$$=
\dfrac{1}{2\pi}\int\limits_{-\infty}^{\infty}e^{ikx_1}
\dfrac{1-i(1-\gamma)k\mu+\gamma(k\mu)^2}{1+(k\mu)^2}E(k)\,dk.
$$

Similarly arguing, we receive, that for function $h_c^-(x_1, \mu) $
precisely same formula takes place
$$
h_c^-(x_1,\mu)=\dfrac{1}{2\pi}
\int\limits_{-\infty}^{\infty}e^{ikx_1}
\dfrac{1+i\gamma k \mu}{1+ik\mu}E(k)\,dk=$$$$=
\dfrac{1}{2\pi}\int\limits_{-\infty}^{\infty}e^{ikx_1}
\dfrac{1-i(1-\gamma)k\mu+\gamma(k\mu)^2}{1+(k\mu)^2}E(k)\,dk.
$$

Therefore
$$
h_c^{\pm}(x_1,\mu)=\dfrac{1}{2\pi}
\int\limits_{-\infty}^{\infty}e^{ikx_1}
\dfrac{1+i\gamma k \mu}{1+ik\mu}E(k)\,dk=$$$$=
\dfrac{1}{2\pi}\int\limits_{-\infty}^{\infty}e^{ikx_1}
\dfrac{1-i(1-\gamma)k\mu+\gamma(k\mu)^2}{1+(k\mu)^2}E(k)\,dk.
$$

Using parity of function $E (k) $ further it is received
$$
h_c^{\pm}(0,\mu)=
\dfrac{1}{2\pi}\int\limits_{-\infty}^{\infty}
\dfrac{1-i(1-\gamma)k\mu+\gamma(k\mu)^2}{1+(k\mu)^2}E(k)\,dk=
$$
$$
=\dfrac{1}{2\pi}\int\limits_{-\infty}^{\infty}
\dfrac{1+\gamma(k\mu)^2}{1+(k\mu)^2}E(k)\,dk=
\dfrac{1}{\pi}\int\limits_{0}^{\infty}
\dfrac{1+\gamma(k\mu)^2}{1+(k\mu)^2}E(k)\,dk.
\eqno{(3.7)}
$$

\begin{center}
\item{} \section{Characteristic system and Fredholm integral equation}
\end{center}

Now we will substitute Fourier integrals (3.6) and (3.5), and also equality
(3.7) in the equations (3.3) and (3.4). We receive characteristic system
of the equations
$$
\Phi(k,\mu)(1+ik\mu)=E(k)(1+i\gamma k\mu)+
$$
$$
=|\mu|\Bigg[-2qU_{sl}(q)+2(1-\gamma)(2-q)G_v|\mu|
%$$
%$$
-\dfrac{q}{\pi}
\int\limits_{0}^{\infty}\dfrac{1+\gamma(k\mu)^2}{1+(k\mu)^2}E(k)\,dk\Bigg],
\eqno{(4.1)}
$$
$$
E(k)=\dfrac{1}{\sqrt{\pi}}\int\limits_{-\infty}^{\infty}
e^{-t^2}\Phi(k,t)dt.
\eqno{(4.2)}
$$

From equation (4.1) we receive
$$
\Phi(k,\mu)=\dfrac{1+i\gamma k\mu}{1+ik\mu}E(k)+
$$
$$+\dfrac{|\mu|}{1+ik\mu}\Bigg[-2qU_{sl}(q)+
2(1-\gamma)(2-q)G_v|\mu|-
$$
$$
-\dfrac{q}{\pi}
\int\limits_{0}^{\infty}\dfrac{1+\gamma(k\mu)^2}{1+(k\mu)^2}E(k)\,dk\Bigg],
\eqno{(4.3)}
$$

Let us substitute $ \Phi(k,\mu) $, defined by equality (4.3), in (4.2).
We receive, that:
$$
E(k)L(k)=-2qU_{sl}(q)T_1(k)+2(1-\gamma)(2-q)G_v T_2(k)-
$$
$$
-\dfrac{q}{\pi^{3/2}}\int\limits_{0}^{\infty}
E(k_1)dk_1\int\limits_{-\infty}^{\infty}
\dfrac{e^{-t^2}|t|(1-ikt)(1+\gamma k^2t^2)dt}{(1+k^2t^2)(1+k_1^2t^2)}.
\eqno{(4.4)}
$$
Here
$$
T_n(k)=\dfrac{2}{\sqrt{\pi}}\int\limits_{0}^{\infty}
\dfrac{e^{-t^2}t^n\,dt}{1+k^2t^2},\qquad n=0,1,2,\cdots,
$$
$L(k)$ is the dispersion function,

\begin{figure}[h]
\begin{center}
\includegraphics[width=15.0 cm, height=10 cm]{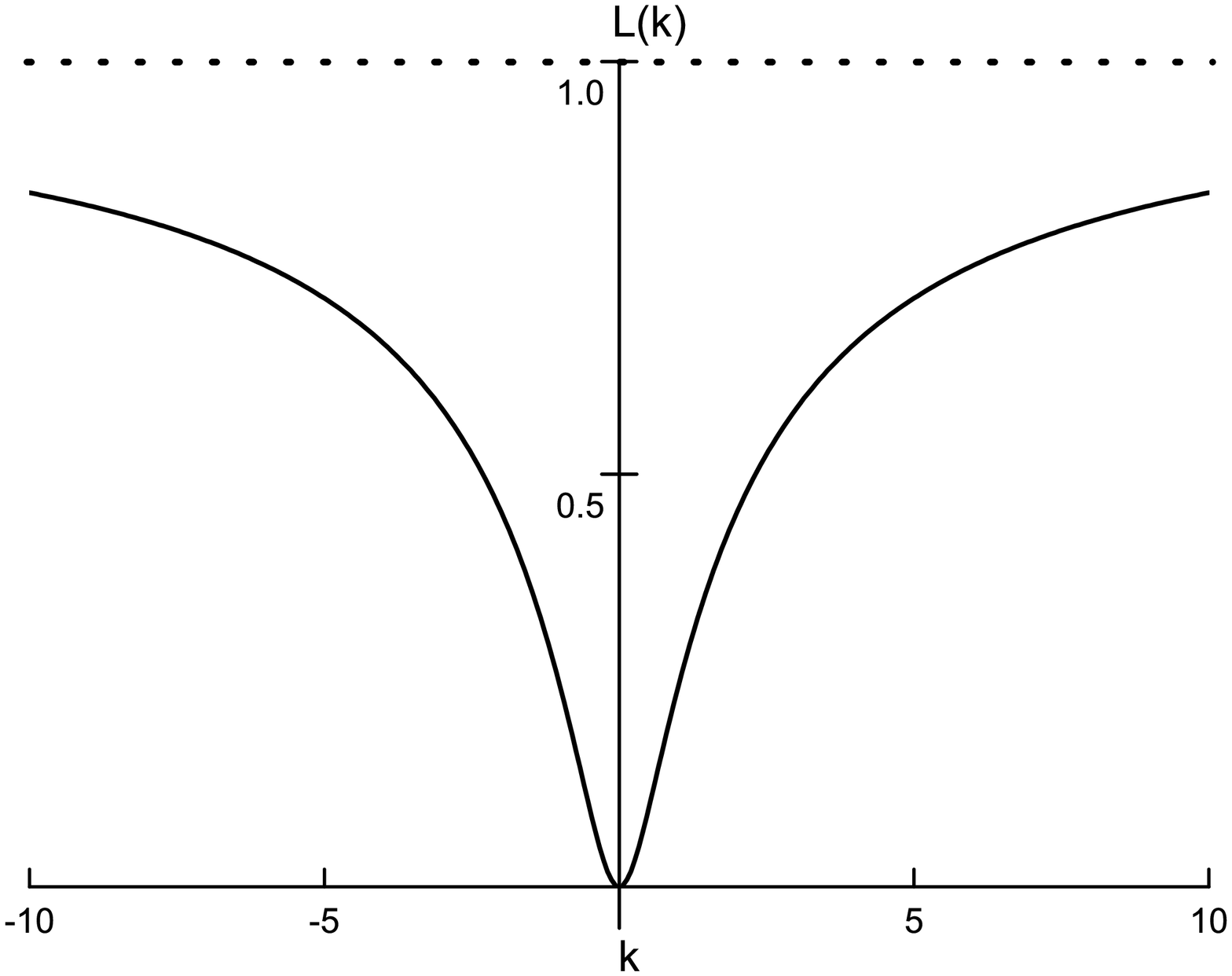}
\end{center}
\begin{center}
{Fig. 1. Dispersion function $L(k)$ at $\gamma=0$.}
\end{center}
\end{figure}

\begin{figure}[h]
\begin{center}
\includegraphics[width=15.0 cm, height=10 cm]{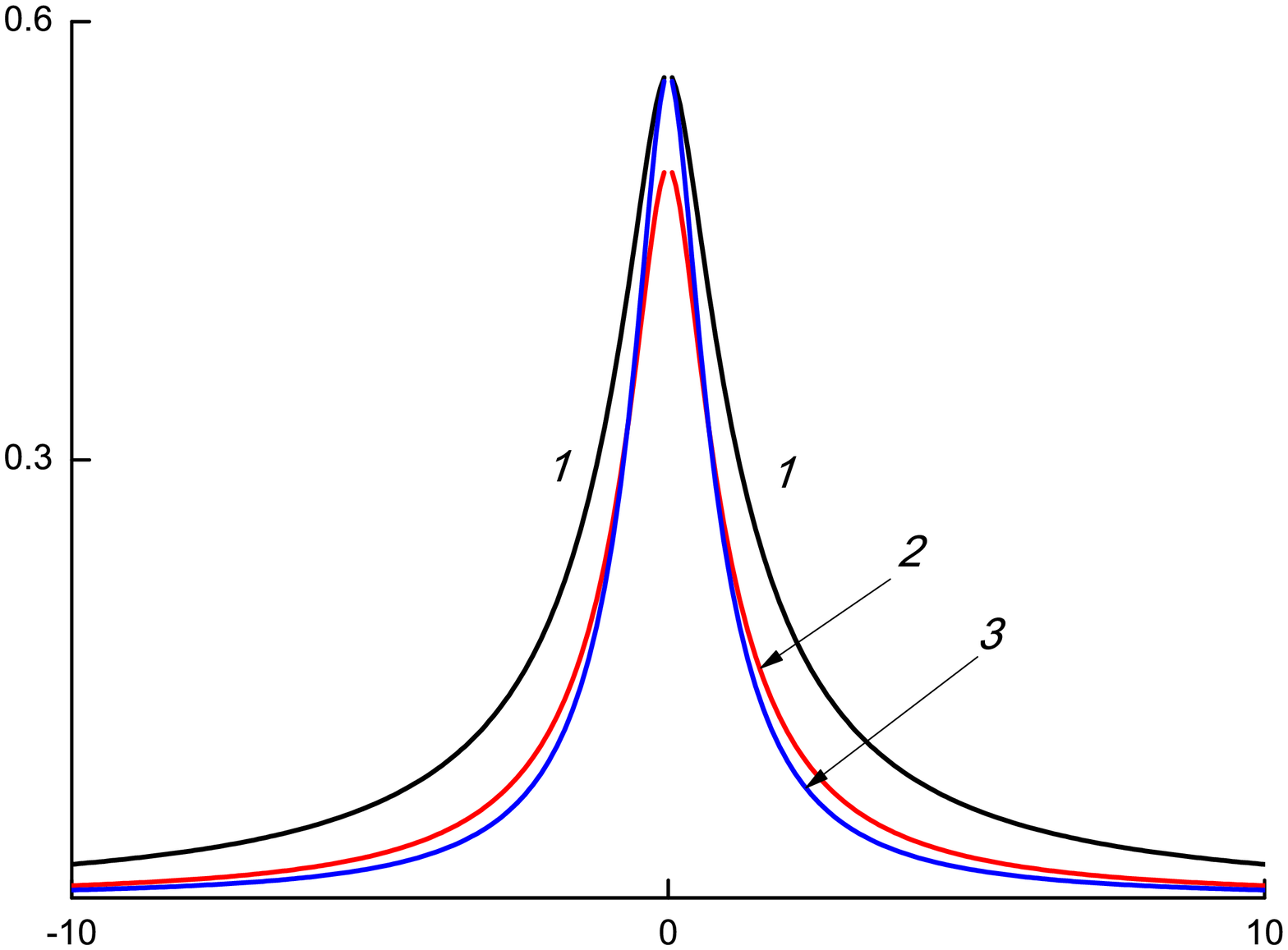}
\end{center}
\begin{center}
{Fig. 2. Integrals type $T_n(x)$ at $n=1,2,3$ (curves $1,2,3$ accordingly).}
\end{center}
\end{figure}

%\clearpage

$$
L(k)=1-\dfrac{1}{\sqrt{\pi}}\int\limits_{-\infty}^{\infty}
e^{-t^2} \dfrac{1-i(1-\gamma)kt+\gamma (kt)^2}{1+(kt)^2}dt=
$$
$$
=1-\dfrac{2}{\sqrt{\pi}}\int\limits_{0}^{\infty}
e^{-t^2}\dfrac{1+\gamma (kt)^2}{1+(kt)^2}dt=
$$
$$
=1-T_0(k)-\gamma k^2T_2(k)=(1-\gamma)k^2T_2(k).
$$
It is easy to see, that
$$
1-T_0(k)=1-\dfrac{1}{\sqrt{\pi}}
\int\limits_{-\infty}^{\infty}\dfrac{e^{-t^2}dt}{1+k^2t^2}=%$$$$=
1-\dfrac{2}{\sqrt{\pi}}\int\limits_{0}^{\infty}\dfrac{e^{-t^2}dt}
{1+k^2t^2}=k^2 T_2(k).
$$
Besides, internal integral in (4.4) we will transform
also we will designate as follows
$$
J(k,k_1)=\dfrac{1}{\sqrt{\pi}}\int\limits_{-\infty}^{\infty}
\dfrac{e^{-t^2}|t|(1-ikt)(1+\gamma k^2t^2)dt}{(1+k^2t^2)(1+k_1^2t^2)}=
$$
$$
=\dfrac{2}{\sqrt{\pi}}\int\limits_{0}^{\infty}
\dfrac{e^{-t^2}t(1+\gamma k_1^2t^2)\,dt}
{(1+k^2t^2)(1+k_1^2t^2)}.
$$

Further  more general integrals will be necessary for us
$$
J^{(m)}(k,k_1)=\dfrac{2}{\sqrt{\pi}}\int\limits_{0}^{\infty}
\dfrac{e^{-t^2}t^m(1+\gamma k_1^2t^2)\,dt}
{(1+k^2t^2)(1+k_1^2t^2)}, \qquad m=1,2,\cdots,
$$
besides
$$
J^{(1)}(k,k_1)=J(k,k_1).
$$

Let us copy now the equation (4.4) by means of the previous equality in
the following form
$$
E(k)L(k)+\dfrac{q}{\pi}\int\limits_{0}^{\infty} J(k,k_1)E(k_1)\,dk_1=
$$$$=
-2qU_{sl}(q)T_1(k)+2(1-\gamma)(2-q)G_v T_2(k).
\eqno{(4.5)}
$$

The equation (4.5) is integral equation Фредгольма of the second
kind. The kernel of this integral equation is the sum
$$
J(k,k_1)=J_1(k,k_1)+\gamma k^2J_3(k,k_1),
$$
where
$$
J_n(k,k_1)=\dfrac{2}{\sqrt{\pi}}\int\limits_{0}^{\infty}
\dfrac{e^{-t^2}t^n\,dt}{(1+k^2t^2)(1+k_1^2t^2)},
$$
besides
$$
J_1(0,k_1)=T_1(k_1), \qquad J_1(k,0)=T_1(k).
$$

We note that
$$
T_n(k)=T_n(0)-k^2T_{n+2}(k), \qquad T_n(0)=\dfrac{2}{\sqrt{\pi}}
\int\limits_{0}^{\infty}e^{-t^2}t^ndt,
$$
besides
$$
T_0(0)=1,\qquad T_1(0)=\dfrac{1}{\sqrt{\pi}}, \qquad T_2(0)=\dfrac{1}{2},\qquad
T_3(0)=\dfrac{1}{\sqrt{\pi}},
$$
$$
T_4(0)=\dfrac{3}{4},\qquad T_5(0)=\dfrac{2}{\sqrt{\pi}},\quad  \cdots.
$$

\begin{center}
\item{} \section{Neumann series and infinitely system of equations}
\end{center}

Further we also search for the solution in the form of series.
These series are Neumann's series. Members of these series are
expressed by multiple integrals.
Multiplicity of integrals grows with growth of number of the member of the
series.

Let us consider, that the gradient of mass velocity in the equation (4.5) is set.
We will expand solutions of the characteristic
systems (4.3) and (4.5) abreast on degrees
diffusion coefficient $q $
$$
E(k)=G_v(1-\gamma)(2-q)\Big[E_0(k)+q\,E_1(k)+q^2\,E_2(k)+\cdots\big],
\eqno{(5.1)}
$$
$$
\Phi(k,\mu)=G_v(1-\gamma)(2-q)\Big[
\Phi_0(k,\mu)+q\Phi_1(k,\mu)+q^2\Phi_2(k,\mu)+\cdots\Big].
\eqno{(5.2)}
$$

For velocity of sliding $U_{sl}(q)$ we will search thus in the form
$$
U_{sl}(q)=G_v(1-\gamma)\dfrac{2-q}{q}
\Big[U_0+U_1q+U_2q^2+\cdots+U_nq^n+\cdots\Big].
\eqno{(5.3)}
$$

Expansion (5.1) and (5.2) according to (3.5) and (3.6) mean, that
mass velocity and the function of distribution corresponding
to continuous spectrum, are in the form of expansion
$$
U_c(x_1)=G_v(1-\gamma)(2-q)\dfrac{1}{2\pi}\int\limits_{-\infty}^{\infty}
e^{ikx_1}\Big[E_0(k)+qE_1(k)+q^2E_2(k)+\cdots\Big]dk,
$$
$$
h_c(x_1,\mu)=G_v(1-\gamma)(2-q)\dfrac{1}{2\pi}\times $$$$ \times
\int\limits_{-\infty}^{\infty}
e^{ikx_1}\Big[\Phi_0(k,\mu)+q\Phi_1(k,\mu)+q^2\Phi_2(k,\mu)+\cdots\Big]dk.
$$

The offered method allows to consider expansion in series
(5.1) - (5.3) which are assumed converging,
at all values of coefficient $q: 0\leqslant q
\leqslant 1$. Besides, it is necessary to notice, that a developed method
it is possible to apply not only to problems with mirror--diffusion
boundary conditions. The method suits and those classes of problems,
for example, with diffusion boundary conditions ($q=1$), which
it is impossible to solve analytically.

Let us substitute numbers (5.1) -- (5.3) in the equations (4.3) and (4.5).
We receive system of the equations
$$
\Big[E_0(k)+qE_1(k)+q^2E_2(k)+\cdots\Big]L(k)=
$$
$$
=T_2(k)-\Big[U_0+qU_1+q^2U_2+\cdots\Big]T_1(k)-
$$
$$
-q\cdot\dfrac{1}{\pi}\int\limits_{0}^{\infty}J(k,k_1)
\Big[E_0(k_1)+qE_1(k_1)+q^2E_2(k_1)+\cdots\Big]\,dk_1
$$
and
$$
\Phi_0(k,\mu)+q\Phi_1(k,\mu)+q^2\Phi_2(k,\mu)+\cdots= \hspace{5cm}
$$
$$
\hspace{4cm}=\dfrac{1+i\gamma k\mu}{1+ik\mu}
\Big[E_0(k)+qE_1(k)+q^2E_2(k)+\cdots\Big]+
$$
$$
+\dfrac{|\mu|}{1+ik\mu}\Big[|\mu|-U_0+qU_1+q^2U_2+\cdots\Big]-
$$
$$
-q\cdot\dfrac{1}{\pi}\int\limits_{0}^{\infty}
\dfrac{1+i\gamma k_1\mu}{1+ik_1\mu}\Big[E_0(k_1)+qE_1(k_1)+
q^2E_2(k_1)+\cdots\Big]\,dk_1.
$$

Now these integral equations of
break up to equivalent infinite system of the equations.
In zero approach it is received the following system of the equations
$$
E_0(k)L(k)=T_2(k)-U_0T_1(k),
\eqno{(5.4)}
$$
$$
\Phi_0(k,\mu)=\dfrac{1+i\gamma k\mu}{1+ik\mu}E_0(k)+
\dfrac{|\mu|}{1+ik\mu}\Big[|\mu|-U_0\Big],
\eqno{(5.5)}
$$

In first approximation we have
$$
E_1(k)L(k)=-U_1T_1(k)-\dfrac{1}{\pi}\int\limits_{0}^{\infty}
J(k,k_1)E_0(k_1)dk_1,
\eqno{(5.6)}
$$
$$
\Phi_1(k,\mu)=\dfrac{1+i\gamma k\mu}{1+ik\mu}E_1(k)+
$$$$
+\dfrac{|\mu|}{1+ik\mu}\Bigg[-U_1-\dfrac{1}{\pi}
\int\limits_{0}^{\infty}
\dfrac{1+\gamma(k_1\mu)^2}{1+k_1^2\mu^2}E_0(k_1)dk_1\Bigg].
\eqno{(5.7)}
$$

In second approximation we have
$$
E_2(k)L(k)=
-U_2T_1(k)-\dfrac{1}{\pi}\int\limits_{0}^{\infty}J(k,k_1)E_1(k_1)\,dk_1,
\eqno{(5.8)}
$$
$$
\Phi_2(k,\mu)=\dfrac{1+i\gamma k\mu}{1+ik\mu}E_2(k)+
$$$$
+\dfrac{|\mu|}{1+ik\mu}\Bigg[-U_2-\dfrac{1}{\pi}
\int\limits_{0}^{\infty}
\dfrac{1+\gamma(k_1\mu)^2}{1+k_1^2\mu^2}E_1(k_1)dk_1\Bigg].
\eqno{(5.9)}
$$

In $n$-th approximation we have
$$
E_n(k)L(k)=-U_nT_1(k)-\dfrac{1}{\pi}
\int\limits_{0}^{\infty}J(k,k_1)E_{n-1}(k_1)dk_1,
\eqno{(5.10)}
$$
$$
\Phi_n(k,\mu)=\dfrac{1+i\gamma k\mu}{1+ik\mu}E_n(k)+
$$$$
+\dfrac{|\mu|}{1+ik\mu}\Bigg[-U_n-\dfrac{1}{\pi}
\int\limits_{0}^{\infty}
\dfrac{1+\gamma(k_1\mu)^2}{1+k_1^2\mu^2}E_{n-1}(k_1)dk_1\Bigg],
\quad n=1,2,3,\cdots.
\eqno{(5.11)}
$$

\begin{center}
  \item{}\section{Zero approximation}
\end{center}

From the formula (5.4) for zero approximation it is found
$$
E_0(k)=\dfrac{T_2(k)-U_0T_1(k)}{L(k)}.
\eqno{(6.1)}
$$

Zero approximation of mass velocity on the basis of (6.1) is equal
$$
U_c^{(0)}(x_1)=G_v(1-\gamma)\dfrac{2-q}{2\pi}\int\limits_{-\infty}^{\infty}
e^{ikx_1}E_0(k)\,dk=
$$
$$
=G_v(1-\gamma)\dfrac{2-q}{2\pi}
\int\limits_{-\infty}^{\infty}
e^{ikx_1}\dfrac{-U_0T_1(k)+T_2(k)}{L(k)}dk.
\eqno{(6.2)}
$$

According to (6.2) we will impose on zero approximation of mass
velocity the requirement: $U_c (+ \infty) =0$. This condition leads to that
subintegral expression from Fourier integral (6.2) in a point $k=0$
finite. Hence, we should eliminate the pole of the second
order in a point $k=0$ at function $E_0 (k) $.
Noticing, that
$$
T_2(0)=\dfrac{1}{2},\;\; T_1(0)=\dfrac{1}{\sqrt{\pi}},
$$
we seek the zero approximation $U_0$:
$$
U_0=\dfrac{T_2(0)}{T_1(0)}=\dfrac{\sqrt{\pi}}{2}\approx 0.8862.
$$

Let us find expression numerator (6.1)
$$
T_2(k)-U_0T_1(k)
=k^2\int\limits_{0}^{\infty}\dfrac{e^{-t^2}t^3}{1+k^2t^2}
\Big(1-\dfrac{2t}{\sqrt{\pi}}\Big)\,dt=
k^2 \Big[\dfrac{\sqrt{\pi}}{2}T_3(k)-T_4(k)\Big].
$$

Therefore we have
$$
T_2(k)-\dfrac{\sqrt{\pi}}{2}T_1(k)=k^2\varphi_0(k),
$$
where
$$
\varphi_0(k)=\int\limits_{0}^{\infty}
\Big(1-\dfrac{2t}{\sqrt{\pi}}\Big)
\dfrac{e^{-t^2}t^3dt}{1+k^2t^2}=\dfrac{\sqrt{\pi}}{2}T_3(k)-T_4(k).
$$

According to (6.1) and (5.5) we have
$$
E_0(k)=\dfrac{\varphi_0(k)}{(1-\gamma)T_2(k)}
$$
and
$$
\Phi_0(k,\mu)=\dfrac{1+i\gamma k\mu}{1+ik\mu}E_0(k)+
\dfrac{|\mu|}{1+ik \mu}(|\mu|-U_0),
$$
and, hence, in zero approximation mass velocity,
corresponding to the continuous spectrum, it is equal
$$
U^{(0)}(x_1)=G_v(1-\gamma)(2-q)\dfrac{1}{2\pi}
\int\limits_{-\infty}^{\infty}
e^{ikx_1}\dfrac{\varphi_0(k)}{T_2(k)}dk.
$$

Corresponding function of distribution is equal
$$
h_c^{(0)}(x_1,\mu)=G_v2(2-q)\times $$$$ \times
\dfrac{1}{2\pi}\int\limits_{-\infty}^{\infty}
\Bigg[\dfrac{1+i\gamma k\mu}{1+ik\mu}\dfrac{\varphi_0(k)}{T_2(k)}+
(1-\gamma)\dfrac{\mu^2-U_0|\mu|}{1+ik\mu}\Bigg]
e^{ikx_1}dk.
$$

\begin{center}
  \item{}\section{First approximation}
\end{center}

Let us pass to the first approximation. As a first approximation from
equations (5.6), (5.7) we find amendments of spectral density
mass velocity and distribution function
$$
E_1(k)=-\dfrac{1}{L(k)}\Big(U_1T_1(k)+\dfrac{1}{1-\gamma}\cdot\dfrac{1}{\pi}
\int\limits_{0}^{\infty}
\dfrac{J(k,k_1)}{T_2(k_1)}\varphi_0(k_1)dk_1\Big)
\eqno{(7.1)}
$$
and
$$
\Phi_1(k,\mu)=\dfrac{1+i\gamma k\mu}{1+ik\mu}E_1(k)-
$$$$-\dfrac{|\mu|}{1+ik\mu}
\Big[U_1+\dfrac{1}{\pi}\int\limits_{0}^{\infty}
\dfrac{1+i\gamma k_1\mu}{1+ik_1\mu}E_0(k_1)dk_1\Big].
\eqno{(7.2)}
$$

Under these amendments we will construct mass velocity and function
of distribution as the first approximation
$$
U_c(x_1)=G_v(1-\gamma)(2-q)\dfrac{1}{2\pi}
\int\limits_{-\infty}^{\infty}e^{ikx_1}\Big[E_0(k)+qE_1(k)\Big]dk
$$
and
$$
h_c(x_1)=G_v(1-\gamma)(2-q)\dfrac{1}{2\pi}
\int\limits_{-\infty}^{\infty}e^{ikx_1}
\Big[\Phi_0(k,\mu)+q\Phi_1(k,\mu)\Big]dk.
$$

The first amendment to mass velocity looks like
$$
U_c^{(1)}(x_1)=G_v(1-\gamma)\dfrac{2-q}{2\pi}\int\limits_{-\infty}^{\infty}
e^{ikx_1}E_1(k)\,dk.
$$

The requirement $U_c(+ \infty) =0$ leads to the finiteness requirement
subintegral expression in the previous Fourier integral.
In expression (7.1) we will substitute decomposition
$$
T_1(k)=\dfrac{1}{\sqrt{\pi}}-k^2T_3(k)
$$
and
$$
J^{(1)}(k,k_1)=J^{(1)}(0,k_1)-k^2J^{(3)}(k,k_1).
$$

It is as a result received, that
$$
E_1(k)=-\dfrac{1}{(1-\gamma)k^2T_2(k)}\Bigg[U_1\dfrac{1}{\sqrt{\pi}}+
\dfrac{1}{1-\gamma}\dfrac{1}{\pi}\int\limits_{0}^{\infty}
J^{(1)}(0,k_1)\dfrac{\varphi_0(k_1)}{T_2(k_1)}dk_1-
$$
$$
-k^2\Bigg(U_1T_3(k)+\dfrac{1}{1-\gamma}\dfrac{1}{\pi}
\int\limits_{0}^{\infty}J^{(3)}(k,k_1)
\dfrac{\varphi_0(k_1)}{T_2(k_1)}dk_1\Bigg)\Bigg].
$$

From this expression it is visible, that for elimination of the pole
of the second order in zero, it is necessary to demand, that
$$
U_1=-\dfrac{1}{1-\gamma}\dfrac{1}{\sqrt{\pi}}\int\limits_{0}^{\infty}
J^{(1)}(0,k_1)\dfrac{\varphi_0(k_1)}{T_2(k_1)}dk_1=
$$
$$
=-\dfrac{1}{1-\gamma}\dfrac{1}{\sqrt{\pi}}\int\limits_{0}^{\infty}
\Big[T_1(k_1)+\gamma k_1^2T_3(k_1)\Big]
\dfrac{\varphi_0(k_1)}{T_2(k_1)}dk_1=
$$
$$
\approx\dfrac{ 0.1405+0.2009\gamma}{1-\gamma}.
$$

Now the spectral density of mass velocity is equal
$$
E_1(k)=\dfrac{1}{(1-\gamma)T_2(k)}\Bigg[U_1T_3(k)+\dfrac{1}{1-\gamma}
\dfrac{1}{\pi}\int\limits_{0}^{\infty}J^{(3)}(k,k_1)
\dfrac{\varphi_0(k_1)}{T_2(k_1)}dk_1\Bigg].
$$

Let us transform expression in a square bracket from
last expression
$$
U_1T_3(k)+\dfrac{1}{1-\gamma}
\dfrac{1}{\pi}\int\limits_{0}^{\infty}J^{(3)}(k,k_1)
\dfrac{\varphi_0(k_1)}{T_2(k_1)}dk_1=
$$
$$
=\dfrac{1}{1-\gamma}\dfrac{1}{\pi}\int\limits_{0}^{\infty}
S(k,k_1)\dfrac{\varphi_0(k_1)}{T_2(k_1)}dk_1.
\eqno{(7.3)}
$$
Here
$$
S(k,k_1)=J^{(3)}(k,k_1)-\sqrt{\pi}T_3(k)J^{(1)}(0,k_1)=
$$
$$
=J_3(k,k_1)-\sqrt{\pi}T_3(k)T_1(k_1)+\gamma k_1^2[J_5(k,k_1)-\sqrt{\pi}
T_3(k)T_3(k_1)]=
$$
$$
=S_1(k,k_1)+\gamma k_1^2S_2(k,k_1),
$$
where
$$
S_1(k,k_1)=J_3(k,k_1)-\sqrt{\pi}T_3(k)T_1(k_1),
$$
$$
S_2(k,k_1)=J_5(k,k_1)-\sqrt{\pi}T_3(k)T_3(k_1).
$$

Now the spectral density is equal
$$
E_1(k_1)=\dfrac{\varphi_1(k_1)}{(1-\gamma)^2T_2(k_1)}.
$$
Here the designation is entered
$$
\varphi_1(k_1)=\dfrac{1}{\pi}\int\limits_{0}^{\infty}
S(k_1,k_2)\dfrac{\varphi_0(k_2)}{T_2(k_2)}\,dk_2.
$$

Now spectral density (7.1) and (7.2) completely
are constructed. Velocityof sliding is as the first approximation equal
$$
U_{sl}(q)=G_v(1-\gamma)\dfrac{2-q}{q}
\Big[0.8862+\dfrac{0.1405+0.2009\gamma}{1-\gamma}q\Big]=
$$
$$
=G_v\dfrac{2-q}{q}\Big[0.8862(1-\gamma)+(0.1405+0.2009\gamma)q\Big].
$$

\begin{center}
  \item{}\section{Second approximation}
\end{center}

Let us pass to the second approximation of the problem.
We take the equations (5.8) and (5.9).

From the equation (5.8) it is found
$$
E_2(k)=-\dfrac{1}{L(k)}\Big[U_2T_1(k)+\dfrac{1}{\pi}
\int\limits_{0}^{\infty}
J(k,k_1)E_1(k_1)\,dk_1\Big].
\eqno{(8.1)}
$$

The second amendment to mass speed looks like
$$
U_c^{(2)}(x_1)=G_v(1-\gamma)\dfrac{2-q}{2\pi}\int\limits_{-\infty}^{\infty}
e^{ikx_1}E_2(k)\,dk.
$$

The condition $U_c(+ \infty) =0$ leads to the requirement of
limitation of function $E_2(k) $ in a point $k=0$.
Let us eliminate the pole of the second order in the point $k=0$
in the right part of equalities for $E_2(k)$.
For this purpose we will present expression in a square bracket from (8.1) in
the form
$$
E_2(k)=-\dfrac{1}{(1-\gamma)k^2T_2(k)}
\Bigg[U_2\Big(\dfrac{1}{\sqrt{\pi}}-k^2T_3(k)\Big)+
$$
$$
+\dfrac{1}{(1-\gamma)^2}\dfrac{1}{\pi}\int\limits_{0}^{\infty}
\Big[J^{(1)}(0,k_1)-k^2J^{(3)}(k,k_1)\Big]
\dfrac{\varphi_1(k_1)}{T_2(k_1)}\Bigg].
$$

From this equality it is visible, that for pole elimination in zero
it is necessary to demand, that
$$
U_2=-\dfrac{1}{(1-\gamma)^2}\dfrac{1}{\sqrt{\pi}}
\int\limits_{0}^{\infty}J^{(1)}(0,k_1)\dfrac{\varphi_1(k_1)}{T_2(k_1)}dk_1.
\eqno{(8.2)}
$$

Let us present expression (8.2) in the form convenient for calculations
$$
U_2=-\dfrac{1}{(1-\gamma)^2}\dfrac{1}{\sqrt{\pi}}
\int\limits_{0}^{\infty}\int\limits_{0}^{\infty}\Bigg\{T_1(k_1)S_1(k_1,k_2)+
\gamma\Big[k_1^2T_3(k_1)S_1(k_1,k_2)+$$$$+k_2^2T_1(k_1)S_2(k_1,k_2)\Big]+
\gamma^2k_1^2k_2^2T_3(k_1)S_2(k_1,k_2)\Bigg\}\dfrac{\varphi_0(k_k)dk_1dk_2}
{T_2(k_1)T_2(k_2)}=
$$
$$
=-\dfrac{J_0+\gamma J_1+\gamma^2J_2}{(1-\gamma)^2}.
$$

Here
$$
J_0=\dfrac{1}{\pi^{3/2}}\int\limits_{0}^{\infty}
\int\limits_{0}^{\infty}T_1(k_1)S_1(k_1,k_2)
\dfrac{\varphi_0(k_2)dk_1dk_2}
{T_2(k_1)T_2(k_2)}=0.0116, %(0.01155)
$$
$$
J_1=\dfrac{1}{\pi^{3/2}}\int\limits_{0}^{\infty}\int\limits_{0}^{\infty}
\Big[k_1^2T_3(k_1)S_1(k_1,k_2)+%$$$$+
k_2^2T_1(k_1)S_2(k_1,k_2)\Big]\times$$$$\times
\dfrac{\varphi_0(k_2)dk_1dk_2}
{T_2(k_1)T_2(k_2)}=0.0125, %0.012478
$$
$$
J_2=\dfrac{1}{\pi^{3/2}}\int\limits_{0}^{\infty}\int\limits_{0}^{\infty}
k_1^2k_2^2T_3(k_1)S_2(k_1,k_2)\dfrac{\varphi_0(k_2)dk_1dk_2}
{T_2(k_1)T_2(k_2)}=-0.0306. %-0.030558
$$

Thus, in the second approximation it is found, that
$$
U_2=-\dfrac{J_0+\gamma J_1+\gamma^2 J_2}{(1-\gamma)^2}=
-\dfrac{0.0116+0.0125\gamma -0.0306\gamma^2}{(1-\gamma)^2}.
$$

On the basis of (8.2) from the previous it is found
$$
E_2(k)=-\dfrac{1}{(1-\gamma)T_2(k)}\Bigg[U_2T_2(k)+
\dfrac{1}{(1-\gamma)^2}\dfrac{1}{\pi}\int\limits_{0}^{\infty}
J^{(3)}(k,k_1)\dfrac{\varphi_1(k_1)}{T_2(k_1)}dk_1\Bigg]=
$$
$$
=\dfrac{1}{(1-\gamma)^3T_2(k)}\dfrac{1}{\pi}\int\limits_{0}^{\infty}
\Big[J^{(3)}(k,k_1)-\sqrt{\pi}T_3(k)J^{(1)}(0,k_1)\Big]
\dfrac{\varphi_1(k_1)}{T_2(k_1)}dk_1=
$$
$$
=\dfrac{\varphi_2(k)}{(1-\gamma)^3T_2(k)}.
$$
Here
$$
\varphi_2(k)=\dfrac{1}{\pi}\int\limits_{0}^{\infty}
S(k,k_1)\dfrac{\varphi_1(k_1)}{T_2(k_1)}dk_1,
$$
where
$$
S(k,k_1)=J^{(3)}(k,k_1)-\sqrt{\pi}T_3(k)J^{(1)}(0,k_1).
$$

Under the found amendments of the second order it is possible
to construct the mass velocity and distribution function in the
second approximation
$$
U_c(x_1)=G_v(1-\gamma)(2-q)\dfrac{1}{2\pi}
\int\limits_{-\infty}^{\infty}e^{ikx_1}\Big[E_0(k)+qE_1(k)+q^2E_2(k)\Big]dk
$$
and
$$
h_c(x_1)=G_v(1-\gamma)(2-q)\dfrac{1}{2\pi}
\int\limits_{-\infty}^{\infty}e^{ikx_1}
\Big[\Phi_0(k,\mu)+q\Phi_1(k,\mu)+q^2\Phi_2(k,\mu)\Big]dk.
$$

The amendment $ \Phi_2 (k, \mu)$ we seek from the equation (5.9) with
use of the constructed function $E_2(k)$ and constants $U_2$.

Velocity of sliding is equal in the second approximatiom
$$
U_{sl}(q)=G_v(1-\gamma)\dfrac{2-q}{q}\Big[U_0+U_1q+U_2q^2\Big].
$$

The formula (8.2) we will transform through calculation of the double
integral of the following form
$$
U_2=-\dfrac{1}{(1-\gamma)^2}\dfrac{1}{\pi^{3/2}}
\int\limits_{0}^{\infty}\int\limits_{0}^{\infty}
\dfrac{J^{(1)}(0,k_1)S(k_1,k_2)}{T_2(k_1)T_2(k_2)}\varphi_0(k_2)\,
dk_1dk_2.
$$

Also we will transform the formula for calculation
$\varphi_2(k)$
$$
\varphi_2(k)=\dfrac{1}{\pi^2}
\int\limits_{0}^{\infty}
\int\limits_{0}^{\infty} \dfrac{S(k,k_1)S(k_1,k_2)}{T_2(k_1)T_2(k_2)}
\varphi_0(k_2)dk_1dk_2.
$$

Spending similar reasonings, for $n $-th approximation
($n=1,2, \cdots $)
according to (5.10) and (5.11) it is received
$$\displaystyle
U_n=-\dfrac{1}{(1-\gamma)^n}\dfrac{1}{\sqrt{\pi}}
\int\limits_{0}^{\infty}J^{(1)}(0,k_1)\dfrac{\varphi_n(k_1)}{T_2(k_1)},
$$
besides
$$
E_n(k)=\dfrac{1}{(1-\gamma)^{n+1}}\dfrac{\varphi_n(k)}{T_2(k)},
$$
where
$$
\varphi_{n}(k)=-\dfrac{1}{\pi}\int\limits_{0}^{\infty}S(k,k_1)
\dfrac{\varphi_{n-1}(k_1)}{T_2(k_1)}dk_1,\quad n=1,2,\cdots.
$$

Let us write out $n $th  approximation $U_n $ and $ \varphi_n (k) $,
expressed through multiple integrals and
zero approximation of spectral density of mass velocity
$E_0 (k) $. We have
$$
U_n=-\dfrac{1}{(1-\gamma)^n\pi^{n-1/2}}\int\limits_{0}^{\infty}\cdots
\int\limits_{0}^{\infty}J^{(1)}(0,k_1)\dfrac{S(k_1,k_2)\cdots
S(k_{n-1},k_n)}{T_2(k_1)\cdots T_2(k_n)}
\times
$$
$$
\times \varphi_0(k_n)\,dk_1\cdots dk_n,
$$

$$
E_n(k)=\dfrac{1}{(1-\gamma)^{n+1}\pi^{n}T_2(k)}\int\limits_{0}^{\infty}
\cdots \int\limits_{0}^{\infty}\dfrac{S(k,k_1)S(k_1,k_2)\cdots
S(k_{n-1},k_n)}{T_2(k_1)\cdots T_2(k_n)}
\times $$$$ \times
\varphi_0(k_n)dk_1\cdots dk_n,
$$

$$
\varphi_n(k)=\dfrac{1}{\pi^{n}}\int\limits_{0}^{\infty}\cdots
\int\limits_{0}^{\infty}\dfrac{S(k,k_1)S(k_1,k_2)\cdots S(k_{n-1},k_n)}
{T_2(k_1)\cdots T_2(k_n)}
\times$$$$\times
\varphi_0(k_n)dk_1\cdots dk_n,\;\quad n=1,2,\cdots.
$$

\begin{center}
  \item{}\section{Comparison with previous results}
\end{center}

Let us compare zero, the first and the second approximation of
velocity of sliding with the exact solution of the equation (2.11).
It has been above found, that
zero, first and second amendments (coefficients of expansion)
of velocity of sliding are accordingly equal
$$
U_0=\dfrac{\sqrt{\pi}}{2}=0.8862,
$$
$$
U_1=\dfrac{0.1405+0.2009 \gamma}{1-\gamma},
$$
$$
U_2=-\dfrac{0.0116+0.0125\gamma -0.0306\gamma^2}{(1-\gamma)^2}.
$$

On the basis of these equalities we will write out dimensionless
velocity of sliding of moderately dense gas in the second approach
$$
U_{sl}(q)=G_v(1-\gamma)\dfrac{2-q}{q}\Big[0.8862+
\dfrac{0.1405+0.2009 \gamma}{1-\gamma}q-$$$$
-\dfrac{0.0116+0.0125\gamma -0.0306\gamma^2}{(1-\gamma)^2}q^2\Big].
\eqno{(9.1)}
$$

We will consider the first limiting case, when $q\to 0$, i.e.
we consider the Kramers' problem with the pure diffusion boundary
conditions. From equality (9.1) considering, that parametre $ \gamma $
satisfies to an inequality: $ \gamma \ll 1$, we find, that
$$
U_{sl}(q)=G_v(1.0152-0.6862\gamma).
\eqno{(9.2)}
$$

Из результатов работы \cite{method2012} вытекает, что точное
выражение безразмерной скорости скольжения равно:
$$
U_{sl}(q)=G_v(V_1-\dfrac{1}{\sqrt{2}})=G_v(1.0162-0.7071\gamma).
\eqno{(9.3)}
$$

From comparison of formulas (9.2) and (9.3) follows, that for the rarefied
gases ($ \gamma=0$) exact velocity of sliding differs on $0.1 \% $ from
the second approximation of this velocity, and the parametres proportional
to density coefficients $ \gamma $, differ on $2 \% $ in the same
approximation.

Let us consider the second limiting case when moderately dense gas passes
in rarefied, i.e. the case $ \gamma\to 0$. In this case
according to the reasonings giving above velocity of sliding
it is equal
$$
U_{sl}(q)=G_v \dfrac{2-q}{q}\Big[U_0+U_1q+U_2q^2\Big],
$$
or, according to equality (9.1)
$$
U_{sl}(q)=G_v
\dfrac{2-q}{q}\Big[0.8862+0.1405q-0.0116q^2\Big].
\eqno{(9.2)}
$$

According to results from \cite{method2012} velocity of sliding
in the second approximation {\bf in accuracy coincides with expression
(9.2)}. If to take here $q=1$, how it was already specified,
velocity of sliding is equal $U_{sl}(q) =1.0152G_v $, i.e. velocity
slidings in the second approximation differs from the exact
$U_{sl}(q)=V_1G_v=1.0162G_v$ на $0.1\%$.

Let us reduce the formula for velocity of sliding (4.3) in a dimensional form.
Let us write down this formula in the form
$$
u_{sl}=\dfrac{2-q}{q}U(q)\dfrac{lg_v}{l\sqrt{\beta}\nu},
$$
where
$$
U(q)=U_0+U_1q+U_2q^n+\cdots.
$$

Let us choose length of free mean path $l $  agree Cercignani
\cite{Cerc62}: $l =\eta \sqrt{\pi \beta}/\rho $, where $ \eta $
is the dynamic viscosity of gas, $ \rho $ is the its density. For the given
problems $ \eta =\rho/(2\nu \beta) $. Hence, in the dimensional form
velocity of isothermal sliding it is equal
$$
u_{sl}=K_v(q)l\left( \dfrac{du_y(x)}{dx}\right)_{x\to+ \infty},
$$
where $K_v(q)$ is the coefficient of isotermal sliding,
$$
K_v(q)=\dfrac{2-q}{q}U(q)\dfrac{2}{\sqrt{\pi}}.
$$

\begin{center}
\item{}\section{Conclusion}
\end{center}

In this very work developed earlier (see ЖВММФ, 2012, {\bf 52}:3, 539-552)
the new method of solution half-spatial boundary problems of the
kinetic theory is applied
for the solution of the classical problem of the kinetic theory
(Kramers' problem) about isothermal sliding of moderately dense gas
along the flat firm surface. The general mirror--diffusion
boundary conditions (Maxwell conditions) are used. The kinetic
equation with non local integral of collisions is applied.
At the heart of this method the idea to continue distribution function
in the conjugated  half-space $x_1 <0$ lays.
And to include in the kinetic equation
boundary condition in the form of a member of type of a source
on the function of distribution corresponding to the continuous spectrum.
With the help of the Fourier transformation the kinetic equation it is reduced to Fredholm integral equation of the second kind which we solve by 
method of consecutive approximations. An offered method
possesses the quite high efficiency. So, comparison with the exact
solution shows, that in the second approximation an error in  finding
one coefficient does not surpass $0.1 \% $, and the second does not surpass
$2 \% $.

This method has been successfully applied \cite{problem2} and \cite{problem4}
in the solution of some such challenges of the kinetic theory, which do not presume the analytical solution.
%\newpage

\end{document}